\documentstyle[preprint,aps]{revtex}
\newcommand{\bb}{\begin{equation}}
\newcommand{\G}{G\lbrack \chi_2, \chi_1;\psi_2, \psi_1 \rbrack}
\newcommand{\ee}{\end{equation}}

\newcommand{\D}{{\cal D}}
\newcommand{\m}{{1\over 2}}

\newcommand{\Ch}{{\cal H}}
\newcommand{\pe}{\perp}
\newcommand{\dd}{\delta^{'} (\sigma - \sigma^{'})}

\newcommand{\dt}{\delta}
\begin{document}

\preprint{Si/94/06}

\title{A MODEL OF TWO-DIMENSIONAL QUANTUM \\
GRAVITY IN THE STRONG COUPLING REGIME}

\author{J. GAMBOA\thanks{Alexander von Humboldt Fellow}}

\address{Universit\"at-GH-Siegen, D-57068 Siegen, Germany\\
            and \\
Departamento de F\'{\i}sica, Universidad de Santiago de Chile,\\
Casilla 307, Santiago, Chile\thanks{Address after October 17, 1994,
E-mail: jgamboa@lauca.usach.cl}}

\maketitle

\begin{abstract}

A model of two-dimensional quantum gravity that is the analog of the
tensionless string is proposed. The gravitational constant ($k$) is the
analog of the Regge slope ($\alpha^{'}$) and it shows that when
$k\rightarrow \infty$, $2D$ quantum gravity can be understood as a
tensionless string theory embeded in a two-dimensional target space. The
temporal coordinate of the target space play the role of the time and
the wave function can be interpreted as in standard quantum mechanics.

\end{abstract}
\pacs{ PACS numbers: 04.60.+m, 11.17.+y, 97.60.Lf}

The quantization of gravity in comparison with other physical fields is a
problem fundamentally different because in gravity the spacetime itself
evolves dynamically and the concepts of time and wave function seem be very
differents as is usually assumed in quantum mechanics \cite{isham}.

In the last years many efforts has been realized in order to understand
these problems using differents points of view \cite{alvarez}, however
the full problem it is still difficult and it seems to suggest that
simplifications should be introduced in order to reach a more complete
understanding.

In this letter we would like to present a new route in the solution to
these problems proposing a two-dimensional gravity model that is the
analog of string theory in the limit $\alpha^{'} \rightarrow \infty$
({\it i.e.} tensionless string theory). This model can be more easily
interpreted that $2D$ full gravity and could throw some light to these
fundamental problems.

More precisely, we will show below that; $a)$ $k$ is just the analog
of the Regge slope $\alpha^{'}$ and the limit $k \rightarrow \infty$
correspond to the strong coupling regime of gravity, $b)$ $2D$
quantum gravity in the strong coupling limit is exactly a tensionless
string theory embeded in a two-dimensional target space with the
temporal coordinate playing the role of time,
$c)$ The wave function can be interpreted as in conventional
quantum mechanics with the target space considered as genuine
spacetime.

$\underline{{\it The\,\,\, Model}}$

Before to discuss our problem, let us start considering some basic issues
of string theory.

The action of string theory is
\bb
S = -{T\over 2} \int d^2 \sigma \sqrt{-g} g^{\alpha \beta}
\partial_\alpha x^\mu \partial_\beta x^\nu \eta_{\mu \nu}, \label{cuerda}
\ee
where $T=1/\alpha^{'}$ is the tension and it is a parameter that appear in
(\ref{cuerda}) in order to have a dimensionless expression.

Once the equation (\ref{cuerda}) is established one could try to take
the limit $T \rightarrow 0$ (tensionless limit) and to study the
consequences involved, but
infortunately, this action is not well defined and one must modify
(\ref{cuerda}) in order to have a smooth behaviour.

The tensionless limit at the level of the action
is a problem that was studied by Schild many years ago \cite{schild} and
after considered by several other authors \cite{lizzi,bandos,gamboa1,suecos,bn}
in differents contexts.

In essence the procedure delineated above is equivalent
to make the hamiltonian formulation
of (\ref{cuerda}) and to take directly the tensionless limit in the
constraints
\bb
\Ch_\pe = {1\over 2} ( p^2 + T^2 {x^{'}}^2 ),\,\,\,\,\,\,\,\,
\Ch_1 = p x^{'}  \label{ligaduras}
\ee
because in such limit one replace (\ref{ligaduras}) by
\bb
\Ch_\pe = {1\over 2} p^2, \,\,\,\,\,\,\,\,\,
\Ch_1 = p x^{'}   \label{liganulas}
\ee
and the algebra of constraints becomes
\begin{eqnarray}
& \lbrack& \Ch_\pe (\sigma), \Ch_\pe (\sigma^{'}) \rbrack = 0,
\nonumber
\\
&\lbrack& \Ch_\pe (\sigma), \Ch_1 (\sigma^{'}) \rbrack = ( \Ch_\pe (\sigma)
+ \Ch_\pe (\sigma^{'} )) \dd, \label{algebra} \\
& \lbrack& \Ch_1 (\sigma) , \Ch_1 (\sigma^{'}) \rbrack =
( \Ch_1 (\sigma) + \Ch_1 (\sigma^{'})) \dd. \nonumber
\end{eqnarray}

A careful analysis of the action in the tensionless limit reveal that
$g = \det (g_{\alpha \beta})$ is vanishes and, as a consequence, all the
points of the string moves with the velocity of the light.

Thus, one see the constraints of the tensionless string as describing a set
of infinite massless relativistic particles (one for every point
along the string) suplemented by the transversality condition
$p x^{'} = 0$.

Now, we would like to construct an analog model for $2D$ gravity.

Two-dimensional gravity is described by the following action
\bb
 S = -{1\over 2}\int_{\cal M} d^2\sigma \sqrt{-g}
\left[{1\over k} g^{\alpha\beta}\partial_\alpha \phi
\partial_\beta \phi + R\phi +{1\over 4\sqrt{k}}
\mu^2 e^\phi \right], \label{g1}
\ee
where ${\cal M}$ is the manifold where the theory is defined,
$g^{\alpha \beta}$ the two dimensional metric, $R$ the scalar curvature,
$\mu^2$ the cosmological constant and $k$ the gravitational constant.

The constraints derived from (\ref{g1}) are\cite{marnelious,armenios,gamboa1}

\begin{eqnarray}
 \Ch_\pe& =&  {1\over 2} [ - 4k\,\,{(g_{11}\pi^{11})}^2 +
 {1\over k}{\phi^{'}}^2 -
4\sqrt{k}\,\,(g_{11}\pi^{11})P_\phi -
{1\over \sqrt{k}}
{{g^{'}_{11}}\over g_{11}}\phi^{'} + 2\phi^{''} +{1\over 4} \mu^2 g_{11}\,\,
e^{\phi/\sqrt{k}}], \nonumber \\
 {\Ch}_1 &=& P_\phi \phi^{'} - 2 g_{11}{\pi^{'}}^{11} - \pi^{11}g^{'}_{11},
\label{a2}
\end{eqnarray}
where $\pi^{\alpha\beta} $ is the canonical momentum associated to
$g_{\alpha \beta}$.

In order to simplify the structure of (\ref{a2}) is more convenient to
perform  a set of three successives canonical transformations namely;
\\
$i)$ We replace in (\ref{a2}) the transformation
\bb
\pi = \pi^{11}g_{11}, \,\,\,\,\, \chi= \ln g_{11}. \label{a3}
\ee
in order to reexpress the non-local term that appears in $\Ch_\pe$,
\\
$b)$ We rescale $P_\phi$ and $\phi$
\bb
b = \sqrt{k} P_\phi, \,\,\,\,\,\, {\tilde \phi} = {\phi\over \sqrt{k}},
\ee
$c)$ The constraints are diagonalized by means
 \begin{eqnarray}
& \psi = {\tilde \phi} - \m \chi, \,\,\,\,\,\,\,\,\,\, b=P_\phi
\nonumber \\
& \chi = \chi, \,\,\,\,\,\,\,\,\,\, P = \pi + \m P_\phi, \label{c1}
\end{eqnarray}
where $(\chi, P)$ and $(\psi, b)$ are pairs of canonical variables.

Using (\ref{a3})-(\ref{c1}), the constraints are
\begin{eqnarray}
 {\Ch}_\pe &=&
\m \biggl[ b^2 + {1\over k}{\psi^{'}}^2 -4P^2 -
{1\over 4k}{\chi^{'}}^2
+ {2\over k}\psi^{''} + {1\over k}\chi^{''} +
{1\over 4k}\mu^2 e^{[{({\sqrt{k} + 1/2}) + \psi}]/\sqrt{k}} \biggr],
\nonumber \\
 {\Ch}_1 &=& b \psi^{'} + P \chi^{'} - 2P^{'} + b^{'}. \label{a6}
\end{eqnarray}

Taking the \lq tensionless limit\rq $\,\,$
({\it i.e.} $k \rightarrow \infty $) in (\ref{a6}),
one find
\begin{eqnarray}
 \Ch_\pe &=&
\m [ b^2 -4P^2 ],
\nonumber \\
{\Ch}_1 &=& b \psi^{'} + P \chi^{'} - 2P^{'} + b^{'}. \label{a10}
\end{eqnarray}

Thus, the limit $k \rightarrow \infty$ cancel the
self-interactions that comes from of the cosmological term and
(\ref{a10}) satisfy exactly the algebra (\ref{algebra}).
In consequence, these constraints are the analog of the tensionless
string for the $2D$ gravity case.

The action for this gravity model is
\bb
S~=~ \int d^2\sigma~ [b{\dot \psi} + P{\dot \chi} - N\Ch_\pe - N_1 \Ch_1 ],
\label{action}
\ee
and it is invariant under the following transformation
\begin{eqnarray}
& \dt& \psi = \epsilon b + \epsilon_1 \psi^{'} -
\epsilon_1^{'}, \nonumber
\\
&\dt& \chi = -4\epsilon P + {\epsilon}_1 \chi^{'} - 2 {\epsilon}_1^{'},
\nonumber
\\
&\dt& b = - {({\epsilon}_1 b)}^{'} ,
\nonumber
\\
&\dt& P = - {(\epsilon_1 P)}^{'}
\label{transfor},
\\
&\dt& N = {\dot \epsilon } + N^{'} \epsilon_1 -
N \epsilon^{'}_1 + N^{'}\epsilon - N_1 \epsilon^{'},\nonumber
\\
&\dt& N_1 = {\dot {\epsilon}_1} + N_1^{'}\epsilon_1 -
N_1 \epsilon^{'}_1,\nonumber
\end{eqnarray}
provided that
\bb
\epsilon (\tau_2, x) = 0 = \epsilon (\tau_1, x). \label{ep}
\ee
This last equation suggest that \cite{claudio}
\begin{equation}
{\dot N} = 0, \,\,\,\,\,\,\,\,\,\,\,\, N_1 = 0. \label{gauge}
\end{equation}
is a good gauge condition and the equations of motion are easily solved
giving the general solution
\bb
\Phi_A (\sigma, \tau) = \alpha_A \tau + \beta_A,
\ee
where $\Phi_A = (\psi, \chi)$.

So, one see that the solutions of $2D$ gravity in the strong coupling limit
corresponds to an infinite set of free relativistics particles.

$\underline{{\it Quantization}}$

Now, we quantize the model proposed above following the path integral
approach. However, before to write a formula for the propagation amplitude
one should note that in the strong coupling regime
(unless one specify special boundary conditions)
there are not anomalies in the definition
of the functional
measure\footnote{Our previous work \cite{gamboa1} on tensionless
string theories could be an exception, however here alway we will assume that
the Weyl ordering is implicit.}.

Having in account this fact and assuming like-particle
boundaries conditions for our variables
\begin{eqnarray}
\psi_1  &=& \psi (\tau_1, \sigma), \,\,\,\,\,\,\,\,\,\,
\psi_1 = \psi (\tau_2, \sigma) \nonumber \\
\chi_1 &=& \chi (\tau_1, \sigma), \,\,\,\,\,\,\,\,\,\,
\chi_2 = \chi (\tau_2, \sigma), \label{bc1}
\end{eqnarray}
the propagation amplitude in the proper-time gauge becomes
\begin{eqnarray}
& \G~ =~ {\displaystyle \int} \D N \D N_1 \D b \D \psi \D P \D \chi
{}~ \delta [ {\dot N}]~ \delta [N_1] ~det (M) \nonumber
\\ &
\times \exp \biggl[ i
{\displaystyle \int} d^2\sigma \biggl( b {\dot \psi} + P {\dot \chi} -
N \Ch_\pe - N_1 \Ch_1 \biggr) \biggr], \label{l1}
\end{eqnarray}
where the Faddeev-Popov determinant is computed using (\ref{transfor}) and the
result is

\bb
\det (M) = \det \left( \matrix{ \partial^2_\tau + {\dot N_1}^{'} +
N_1^{'} \partial_\tau - {\dot N_1} \partial_\tau - N_1
\partial_\tau \partial_\sigma
\strut &
{\dot N} + N^{'} \partial_\tau - {\dot N}
\partial_\tau - N \partial_\tau \partial_\sigma \cr
0
\strut &
\partial_\tau + N_1^{'} - N_1 \partial_\sigma \cr}\right). \label{fp}
\ee
The explicit calculation of (\ref{fp}) in the proper-time gauge is
straightforward and only contributes with a constant.

After to integrate in $N_1$ and in the canonical momenta, we obtain
\bb
\G ~=
\int_0^\infty \prod_\sigma d N(\sigma) \int \D \psi \D \chi
e^{i \int d^2\sigma {1\over 2N} {({\dot \psi}^2 - {1\over 4}
{{\dot \chi}}^2)}}, \label{free}
\ee
where the integral in $N(\sigma)$ is a consequence of
the gauge choice and their limits have in account the causality principle.

At this point, one might be tempted to interpret the action
(\ref{free}) as describing two massless relativistics particles moving in
$1+0$ dimensions, but the lack of Lorentz invariance say us that this
interpretation is erroneous. The right
interpretation is to observe that (\ref{free}) describes an infinite set of
massless relativistic particles moving in a flat two-dimensional
target space in accordance with the action
\begin{eqnarray}
S &=& \int d^2\sigma
[ {1\over 2N} \eta^{AB} {\dot \Phi}_A {\dot \Phi}_B],\nonumber
\\
&=& \sum_{k=-\infty}^{+\infty} \int d\tau
{}~{1\over 2N_k}\eta^{AB} {\dot \Phi}_{kA}
{\dot \Phi}_{kB}, \label{target}
\end{eqnarray}
where the $2$-vector $\Phi_A$ has components
$(\psi, {1\over 2}\chi)$ and $\eta^{AB}$ is the Minkowski metric.

Once this observation is made, then one argue that the quantum
dynamics naturally takes place on a spacetime where the $2$-vector $\Phi_A$
is the \lq $2$-position\rq $\,\,$ with $\Phi_0 = \psi$ playing the
role of time.

So, the propagation amplitude
\begin{eqnarray}
G[\Phi_2, \Phi_1] &=& \int_0^\infty d N(\sigma) \int \D \Phi_A
e^{ i \int d^2\sigma
{1\over 2N} \eta^{AB} {\dot \Phi}_A {\dot \Phi}_B}, \nonumber
\\
&=& \prod_k^\infty \biggl( \int_0^\infty dN_k \int \D \Phi_{kA}
e^{i \int d \tau {1\over 2N_k}
\eta^{AB} {\dot \Phi}_{kA} {\dot \Phi}_{kB}} \biggr),
\label{free1}
\end{eqnarray}
and the wave function can be
interpreted as in conventional quantum mechanics.

{}From (\ref{free1}) one should note that the basic quantum associated
to the $2D$ quantized gravitational field (\lq gravitons\rq ) are simply
massless relativistic particles moving in a two-dimensional spacetime.
Thus, at this level, $2D$ quantum gravity in the strong coupling limit
and tensionless string theory living in a two-dimensional target space
become the same concept.

The propagator (\ref{free1}) can be computed discretizing (\ref{bc1})
\bb
\Phi_{kA} (\tau_1) = \Phi_{kA1} , \,\,\,\,\,\,\,\,
\Phi_{kA} (\tau_2) = \Phi_{kA2}. \label{bc2}
\ee
As is usual in the quantization of the relativistic particle one can make the
change of variables
\bb
\Phi_{kA} (\tau) = \Phi_{kA} + {\Delta \Phi_{kA}\over \Delta \tau}
(\tau - \tau_1) + \Theta_{kA} (\tau)  \label{solutions}
\ee
where $\Theta_{k,A}$ is a quantum fluctuations that satisfy
\bb
\Theta_{kA} (\tau_2) = 0 = \Theta_{kA} (\tau_1). \label{quant}
\ee

Replacing (\ref{solutions}) in (\ref{free1}) one find
\bb
G~[\Phi_2, \Phi_1] = \prod_k^\infty \biggl( \int_0^\infty dT_k T_k^{-1}
e^{{i\over 2T_k}{(\Delta {\Phi}_k)}^2}\biggr),
\label{parti}
\ee
where $T_k = (\tau_2 - \tau_1)N_k$ and the path integral in
$\Theta$ was computed using (\ref{quant}).

The integrals that appears in (\ref{parti}) are infrared divergents
and one can compute it introducing a massive regulator (plus a Wick rotation),
the final result is
\begin{eqnarray}
G~[\Phi_2, \Phi_1] &=& \prod_k^\infty \biggl(
\ln \vert \Delta \Phi_k \vert \biggr)\nonumber \\
&=& \ln \vert \Delta \Phi \vert
\end{eqnarray}
where $\vert \Delta \Phi \vert = \sqrt{{\Delta \psi}^2 - {1\over 4}
{\Delta \chi}^2}$.

The interactions in this model of gravity only can be understood if the
mechanism that generate dynamically the \lq tension\rq $\,\,$ terms
are known.
So, in this sense, $2D$ quantum gravity in the strong couplig
regime is the analog of an ordinary gauge theory before to break the
gauge symmetry (or probably a topological field theory \cite{witten}).

The results described above also can be extended when the gravity
is coupled to matter fields. In such case the constraints (\ref{a10})
are modified by
\begin{eqnarray}
&\Ch_\pe& \rightarrow \Ch_\pe + \sum_{i=1}^N {1\over 2} P_i^2,
\nonumber \\
&\Ch_1& \rightarrow \Ch_1 + \sum_{i=1}^N P_i f^{'}_i, \label{modi}
\end{eqnarray}
where $f_i$ are $N$ matter fields and $P_i$ their conjugate momenta. The
new terms in (\ref{a10}) are set out in such way that
the algebra (\ref{algebra}) is holds.

When the quantization of (\ref{modi}) is considered, matter and gravity
appears factorized and the full propagator is simply a product of two
propagators.

The reader should note that the \lq graviton\rq $\,\,$ in the model
presented above appears after that the variables were (non-locally)
redefined and the covariant structure of the target space emerged.
This point of view is quite different compared with the standard graviton;
it could be an indication that the spacetime should be replaced by
another concept in a quantum theory of gravity.

It is a pleasure for me to thank Manuel Asorey, Jos\'e Luis Cort\'es and
Lech Szymanowski for fruitful discussions. This work was supported by the
Alexander von Humboldt Foundation.

\end{document}